\documentclass[sigconf,authorversion,nonacm]{acmart}

\usepackage{subcaption}
\usepackage{graphicx}
\usepackage{xcolor,colortbl}
\usepackage{pifont}
\newcommand{\cmark}{\ding{51}}%
\newcommand{\xmark}{\ding{55}}%

\AtBeginDocument{%
  }

\copyrightyear{2025}
\acmYear{2025}
\setcopyright{acmlicensed}\acmConference[ISWC '25]{Proceedings of the 2025 ACM International Symposium on Wearable Computers}{October 12--16, 2025}{Espoo, Finland}
\acmBooktitle{Proceedings of the 2025 ACM International Symposium on Wearable Computers (ISWC '25), October 12--16, 2025, Espoo, Finland}
\acmDOI{10.1145/3715071.3750404}
\acmISBN{979-8-4007-1481-8/2025/10}

\begin{document}

\title{CoughViT: A Self-Supervised Vision Transformer for Cough Audio Representation Learning}


\author{Justin Luong}
\affiliation{%
  \institution{University of New South Wales}
  \city{Sydney}
  \country{Australia}
}
\email{justin.luong@unsw.edu.au}
\orcid{0009-0007-3320-3567}

\author{Hao Xue}
\affiliation{%
  \institution{University of New South Wales}
  \city{Sydney}
  \country{Australia}
}
\email{hao.xue1@unsw.edu.au}
\orcid{0000-0003-1700-9215}

\author{Flora D. Salim}
\affiliation{%
  \institution{University of New South Wales}
  \city{Sydney}
  \country{Australia}
}
\email{flora.salim@unsw.edu.au}
\orcid{0000-0002-1237-1664}

\begin{abstract}
Physicians routinely assess respiratory sounds during the diagnostic process, providing insight into the condition of a patient's airways. In recent years, AI-based diagnostic systems operating on respiratory sounds, have demonstrated success in respiratory disease detection. These systems represent a crucial advancement in early and accessible diagnosis which is essential for timely treatment. However, label and data scarcity remain key challenges, especially for conditions beyond COVID-19, limiting diagnostic performance and reliable evaluation. In this paper, we propose CoughViT, a novel pre-training framework for learning general-purpose cough sound representations, to enhance diagnostic performance in tasks with limited data. To address label scarcity, we employ masked data modelling to train a feature encoder in a self-supervised learning manner. We evaluate our approach against other pre-training strategies on three diagnostically important cough classification tasks. Experimental results show that our representations match or exceed current state-of-the-art supervised audio representations in enhancing performance on downstream tasks.
\end{abstract}

\ccsdesc[500]{Applied computing~Health informatics}
\ccsdesc[500]{Information systems~Data mining}
\ccsdesc[500]{Computing methodologies~Unsupervised learning}

\keywords{Audio Analysis, Cough Classification, Self-Supervised Learning}


\maketitle

\section{Introduction}
The observation of respiratory sounds plays a crucial role in disease diagnosis. To this end, physicians employ a technique called auscultation to listen to a patient's respiratory sounds through a stethoscope \cite{sarkar2015auscultation}. Respiratory diseases cause physical changes that distinctly alter the characteristics of sounds produced, such as introducing abnormalities like wheezing. These sound changes offer insights into airway conditions and help identify diseases like asthma, COVID-19, and chronic obstructive pulmonary disease (COPD) \cite{kim2022coming}.
While quick and non-invasive, auscultation has drawbacks. Firstly, diagnostic accuracy varies considerably among medical practitioners performing auscultation. Specialists like pulmonologists demonstrate significantly higher accuracy compared to other types of medical practitioners \cite{hafke2019accuracy}. This suggests the potential for improved diagnostic consistency and accuracy. Secondly, auscultation's requirement for physical presence limits its use in telehealth. Patients with limited mobility or residing in remote areas often face challenges with in-person visits, making telehealth an essential service \cite{kim2022coming}. 

Researchers are exploring AI for automatic auscultation to overcome these drawbacks. Automated systems could offer consistent diagnoses and better accessibility \cite{gurung2011computerized}. Respiratory sounds, captured through digital stethoscopes or microphones, are used to train machine learning models for diagnostic tasks. These models demonstrate considerable accuracy in disease detection \cite{kim2022coming}, suggesting that automatic auscultation could be a viable alternative when traditional methods are unavailable. Cough audio-based diagnostic systems in particular show considerable potential for telehealth. Coughs may offer richer diagnostic signals as they transmit sound unimpeded through the airways compared to the indirect sounds captured by stethoscopes through the chest wall \cite{porter2019prospective}. The widespread use of mobile phones also facilitates the collection of training data and provides a convenient avenue to deliver diagnostic aid without specialised equipment. Researchers have demonstrated the ability to train classifiers to effectively detect diseases such as Asthma \cite{amrulloh2015cough}, Bronchitis \cite{pahar2021covid}, and COVID-19 \cite{brown2020exploring} using cough audio. Timely treatment of respiratory conditions is crucial for minimising damage, especially in young children and the elderly. AI-based diagnostic systems are an important step in advancing early and accessible diagnosis, enabling prompt treatment.

While cough audio modelling holds significant promise for respiratory disease diagnosis, we identify three critical challenges in current research. \textbf{(1) Data Scarcity:} Research has disproportionately focused on COVID-19 related datasets, while data for other conditions remains scarce. Effective classifiers for a variety of respiratory diseases are essential for differential diagnosis and identifying comorbidities \cite{ijaz2022towards}. Addressing data scarcity for these conditions would improve the applicability of cough audio modelling in the real world. \textbf{(2) Reliance on Labels:} Current research focuses on supervised learning methods, which require high-quality annotations. High-quality, clinically validated annotations are expensive which results in smaller datasets. Crowd-sourced datasets offer larger volumes of data but their label reliability is a concern \cite{coppock2021covid}. This trade-off represents a label scarcity issue, risking models being under-fitted or misled by inconsistent labels. \textbf{(3) Model Rigidity and Adaptability:} The reliance on traditional statistical models, pose another challenge as they are constrained by rigid assumptions and require extensive manual feature engineering. In contrast, deep learning architectures offer promising adaptability, with techniques like self-supervised and transfer learning presenting viable solutions to the key issue of data scarcity. While some deep learning architectures have been explored, the field's rapid expansion has unveiled models that may be more suitable in building versatile diagnostic systems.

To address these limitations, we propose CoughViT, a domain-specific pre-training framework to learn general purpose cough feature representations. When training data is limited, pre-training a model on a different but related dataset is an effective approach to improve performance \cite{devlin2018bert}. Models are often pre-trained on diverse, rigorously curated datasets like Audioset \cite{gemmeke2017audio}, spanning various domains. However, pre-training on specialised datasets, which contain exclusively in-domain data, can lead to more effective feature representations for downstream tasks within that domain \cite{gu2021domain}. Our approach leverages domain-specific pre-training on the abundant COVID-19 data to enhance performance on other cough audio tasks that are limited by data scarcity. To our knowledge, we are the first to propose an approach to learn general purpose cough feature representations. Our method employs self-supervised learning which circumvents the need to collect annotations for training. We find that our self-supervised task of spectrogram reconstruction yields more generalisable feature representations compared to a supervised approach. We also match the performance of state-of-the-art models pre-trained on larger, extensively labelled datasets. Furthermore, we explore the use of a Vision Transformer (ViT) architecture \cite{dosovitskiy2020image} in cough audio modelling. We find that its natural ability to handle varying input lengths is advantageous for cough audio data, which tends not to conform to standard sizes. We evaluate our representations on three valuable diagnostic tasks. First is COVID-19 detection, which assesses our framework's effectiveness for disease diagnosis. Second, wet-or-dry cough classification, a characteristic physicians frequently use for differential diagnosis \cite{swarnkar2013automatic}. Third, cough detection is useful for overnight monitoring to detect acute health deterioration \cite{barata2023nighttime}.
In summary, our contributions are: (1) We address label scarcity challenges through self-supervised learning from unlabelled data, and propose a novel domain-specific pre-training framework to learn general-purpose cough feature representations.
(3) We empirically demonstrate that the Vision Transformer architecture is a capable model for cough classification and identify its unique practicality in cough audio modelling.

\section{Related Work}


Cough audio datasets, such as COVID-19 Sounds \cite{xia2021covid} and COUGHVID \cite{orlandic2021coughvid}, are commonly crowd-sourced. Researchers develop web and mobile apps that allow users to record their respiratory sounds and submit metadata such as disease status and comorbidities. Leveraging datasets like these, a wide variety of statistical and deep learning techniques have been explored to detect COVID-19 from cough audio~\cite{brown2020exploring, imran2020ai4covid,han2022sounds,xue2021exploring}. Additionally, researchers have also trained classifiers to detect diseases such as Asthma \cite{amrulloh2015cough}, Bronchitis \cite{pahar2021covid}, COPD \cite{claxton2021identifying}, Pneumonia \cite{amrulloh2015cough}, and Tuberculosis \cite{alqudaihi2021cough} using cough audio. Beyond disease detection these models have also excelled on adjacent tasks such as Wet-or-Dry cough classification \cite{orlandic2023semi}, Asthma severity classification \cite{swarnkar2021stratifying}, cough detection \cite{miranda2019comparative}, cough counting \cite{birring2008leicester}, and cough segmentation \cite{amrulloh2015automatic}.
Pre-training is a key method in enhancing model performance when training data is limited. VGGish \cite{hershey2017cnn}, a CNN variant, utilised audio tagging to train on Audioset \cite{gemmeke2017audio}, a general audio dataset. This model has been widely explored to enhance performance on COVID-19 detection \cite{brown2020exploring,despotovic2021detection,dang2022exploring}. Other variants of CNN pre-trained on Audioset, such as ResNet-38 and MobileNetv1, have also demonstrated potential in improving COVID-19 detection \cite{casanova2021transfer}. ResNeSt \cite{zhang2022resnest} explored the impact of cross-modality pre-training on COVID-19 detection \cite{pinkas2020sars}. It also investigated Mockingjay \cite{liu2020mockingjay}, a self-supervised Transformer model pre-trained on the Librispeech dataset \cite{panayotov2015librispeech}. Furthermore, \cite{xue2021exploring} examined the application of contrastive learning, a self-supervised technique, using a Transformer model for domain-specific pre-training on Coswara \cite{sharma2020coswara}.

Spectrograms are a two-dimensional visual representation of audio which enables the use of computer vision models on them. Owing to the success of Vision Transformers in the image domain, they have also been explored for audio. The Audio Spectrogram Transformer (AST) \cite{gong2021ast} was the first to explore the use of Vision Transformers on audio spectrograms. A self-supervised variant, Self-Supervised Audio Spectrogram Transformer (SSAST) \cite{gong2022ssast}, found success with learning general audio features using a joint discriminative and generative pre-training task. 
Audio-MAE \cite{huang2022masked}, an adaptation of ViT-MAE \cite{he2022masked} for audio, focused on a more scalable self-supervised task which performed similarly well, for learning audio features. We extend this research by exploring the effectiveness of self-supervised vision transformers to learn cough feature representations.

\section{Method}

\begin{figure*}[h]
  \centering
  \includegraphics[width=0.8\textwidth]{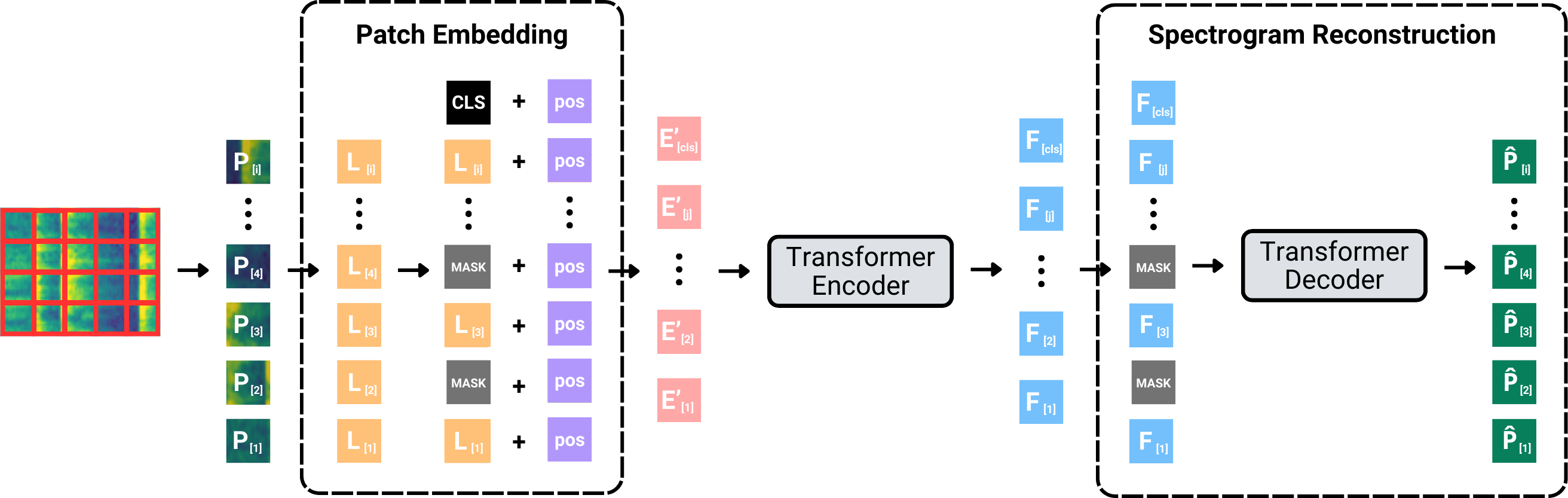}
  \caption{
  Overview of the Vision Transformer architecture for masked data modelling. A random portion of patches are masked and the rest encoded into feature vectors. The decoder then uses these vectors to reconstruct the input spectrogram.
  }
  \Description{
  A diagram illustrating a Vision Transformer architecture for self-supervised pre-training. On the left, an input spectrogram is first divided into a set of patches. Each patch is embedded with positional encoding and classification token pre-pended. A portion of these patch embeddings are masked. The non-masked embeddings pass through a Transformer encoder which produces a set of features. These features are then fed into a Transformer decoder that aims to reconstruct the original spectrogram patches.
  }
  \label{fig:coughvit-architecture}
\end{figure*}

\subsection{Architecture}

In addressing data scarcity in cough classification research, we aim to learn general cough representations to enhance classifier performance across diverse tasks. Accordingly, we focus on Deep Neural Networks (DNN), widely recognised for their effectiveness in representation learning. We also concentrate on architectures compatible with spectrograms, whose compact, time-frequency representation facilitates faster training compared to raw waveforms \cite{purwins2019deep}. Convolutional Neural Networks (CNN) have traditionally been the de facto standard for processing audio spectrograms in deep learning. However, we opt for the Vision Transformer (ViT) architecture for its innate ability to handle varying input lengths. Audio samples rarely conform to standard sizes which complicates adapting pre-trained models to new tasks. A CNN architecture would require altering the input data through methods like padding or utilising more complex architectures. In contrast, ViTs adapt to varied input sizes simply by adjusting their positional encodings and does not require structural changes to the model. This property is advantageous for a pre-trained model framework which may interact with downstream datasets with different input sizes.

\subsubsection{Pre-processing}

An input audio of length $t$ is first resampled to 16000Hz to maintain a consistent scale for Transfer Learning. We transform the audio into a log-mel spectrogram of 128 bins. A frame length $L$ of 25ms and frame hop $H$ of 10ms is used. The number of frames $w$ of the resulting spectrogram can be calculated using $w=\frac{1}{L}(t-H)+1$. For a particular training run, the spectrograms are truncated or padded to a fixed number of frames to enable batching. During pre-training, we do not normalise spectrogram values.

\subsubsection{Patch Generation}
\label{subsubsection:patch-generation}

We first generate a sequence of $16 \times 16$ patches $P$ by dividing the spectrogram. The computational complexity of transformer self-attention layers increases quadratically with the number of patches. Reducing the patch count can play a key part in accelerating training. To that end, we opt for no overlap between patches as it significantly reduces the number of patches to be processed and has been found to be as performant \cite{he2022masked}. 
\begin{equation}
\label{eq:num-patches}
|P| = \left\lfloor \frac{(w - s + \sigma)}{\sigma} \right\rfloor \times \left\lfloor \frac{(h - s + \sigma)}{\sigma} \right\rfloor
\end{equation}
Equation \eqref{eq:num-patches} shows how to calculate the number of patches using the spectrogram dimensions $w \times h$, patch side length $s$, and stride length $\sigma$.

\subsubsection{Patch Embeddings}

These patches are then projected into a set of 768-dimension vectors $L$ using a linear embedding layer.  A learnable classification token is also pre-pended to this set. We use fixed sinusoidal encodings \cite{devlin2018bert}\cite{he2022masked} to retain positional information. The resulting set of positionally aware patch embeddings $E$ is passed into a standard ViT-B encoder \cite{dosovitskiy2020image}. This consists of 12 self-attention layers and produces a set $F$ of 768-dimension encoded patches.

\subsubsection{Masking}
\label{subsubsec:masking}

Masked data modelling, a common self-supervised pre-training task, aims to learn domain knowledge by reconstructing data from masked inputs. In the context of this study, the focus is on reconstructing input spectrograms after masking a portion of their patches. ViT-MAE \cite{he2022masked} and Audio-MAE \cite{huang2022masked} have shown that it is unnecessary to encode the masked patches, as they can be reinserted after the self-attention layers. As explored in section \ref{subsubsection:patch-generation}, this reduction in patch count significantly improves training efficiency. This optimisation is enabled by the transformer encoder's agnosticism to patch count, facilitating a seamless transition to the fine-tuning stage, where masking is not applied. Figure \ref{fig:coughvit-architecture} illustrates the masking process. We randomly mask a portion of patches, resulting in a reduced set of patch embeddings $E'$ representing only unmasked patches. Mask tokens are then inserted after the transformer encoder to restore the patch count and original ordering.

\subsection{Pre-training}
\subsubsection{COVID-19 Sounds}
COVID-19 Sounds is a large crowd-sourced respiratory sound dataset. Each dataset samples consists of a cough, speech, and breathing audio recording. Accompanying metadata such as COVID-19 status, user demographic, and the presence of comorbidities is self-reported by the user. The dataset includes samples from subjects across the globe and spans a wide age range. Of the 53,449 samples 1572 were reported to have had COVID-19 within 14 days of the recording. We use only the cough audio recordings for training in our framework.
We use this dataset for domain-specific pre-training for cough audio modelling. Specifically, the pre-trained model learns to embed audio spectrograms of cough sounds into a latent representation for downstream tasks. 

\subsubsection{Supervised Pre-training}
COVID-19 Sounds provided a user-reported annotation for COVID-19 status. For supervised pre-training, we frame this as a binary classification task and consider the following labels as positive samples: "positiveLast14", "last14", and "yes". We employ no masking in this approach and use the encoded classification token $f_{\text{cls}}$ as the sample-level representation. We employ a simple classifier that predicts the presence of COVID-19. A fully-connected layer maps the encoded token to two output nodes representing the positive and negative classes. A softmax function is then applied to calculate the probability of COVID-19.
However, we identify major limitations with supervised pre-training. First, these methods frequently overfit, generating representations narrowly focused on differentiating the pre-training classes \cite{feng2021rethinking}. Second, their effectiveness relies on accurate labels which is not guaranteed in this case as they are self-reported. Third, this dataset suffers from class imbalance which exacerbates the model's tendency to overfit, limiting its ability to generalise from the pre-training task.

\subsubsection{Self-Supervised Pre-training}

Our self-supervised approach learns only from the audio of each training sample. This circumvents the potentially unreliable labels and class imbalance. Furthermore, by focusing on inherent characteristics of the data, rather than an external label, self-supervised learning encourages more generalisable feature representations. By predicting masked patches, the model learns fundamental visual characteristics of coughs in spectrograms which are broadly applicable in cough classification.

Figure \ref{fig:coughvit-architecture} illustrates our pre-training approach. We randomly mask 75\% of patches. As specified in \ref{subsubsec:masking}, only unmasked patches are encoded. The transformer encoder produces a set of features $F$ which are 768-dimension embeddings representing each unmasked patch. We restore the original patch count and ordering by inserting mask tokens in place of the masked patches. These learnable mask tokens are also 768-dimension vectors.  Similar to the encoder, fixed sinusoidal encodings are added to each vector in this set to encode positional information. These are mapped to 524-dimension vectors through a linear layer. These are then passed through the self-attention layers of the decoder. A final linear layer then outputs a set of 256-dimension vectors $\hat{P}$, representing reconstructions of each original $16 \times 16$ patch.

We use patch normalised mean squared error (MSE) for our loss \cite{he2022masked}. These normalised patches serve at the decoder's prediction targets. Original patches are normalised using its own mean and variance, whereas the total loss in our method  is calculated by taking the MSE for only the masked patches. We also explore a windowed self-attention strategy in the decoder. A standard self-attention layer calculates the relationship of each patch with each other patch. The computational complexity of this operation is hence quadratic to the number of patches. The Swin Transformer \cite{liu2021swin} suggested grouping patches into windows and performing self-attention locally within windows. This approach is much more efficient. Additionally, \cite{huang2022masked} suggested that for audio spectrograms the important relationships are in surrounding patches. In their experiments, they found the use of a windowed self-attention mechanism during self-supervised pre-training to be superior to global and hybrid approaches. Thus, we also explore the use of windowed self-attention in our decoder. We group the patches into $4 \times 4$ non-overlapping windows. We use 16 self-attention layers in the decoder. We also shift between two window configurations to introduce cross-window relationships \cite{liu2021swin,huang2022masked}.

\section{Experiments}

\subsection{Benchmark Datasets}

We evaluate our approach on three distinct cough classification tasks of diagnostic value.
(1) \textbf{COVID-19 Detection:}
We use the Second Dicova Challenge dataset \cite{sharma2022second}, a curated subset of the Coswara  dataset \cite{sharma2020coswara}, to evaluate our models for COVID-19 detection. This dataset is comprised of crowd-sourced audio recordings of cough, speech, and breath, each labelled with self-reported COVID-19 status. Produced for a 2021 research challenge, the dataset's authors excluded ambiguous labels such as "recovered" or "exposed", in addition to recordings that were too brief for analysis. The development dataset comprises data from 965 subjects, 172 of which were reported to have COVID-19 or COVID-19 like symptoms. We adopt the competition's five-fold cross-validation strategy, utilising the provided train-validation splits. We use only each sample's cough recording for classification.
(2) \textbf{Wet-or-Dry Classification:}
The COUGHVID Dataset \cite{orlandic2021coughvid} contains crowd sourced cough audio recordings along with various labels annotated by four expert physicians. We focus on the task of Wet-or-Dry Cough classification. Given the challenges of low inter-rater reliability for this label, highlighted by the original authors, we opt to utilise the annotations provided by expert 4 exclusively.
(3) \textbf{Cough Detection:}
The Edge-AI Cough Counting Dataset \cite{orlandic2023multimodal} is a multi-modal dataset of cough and non-cough sounds with data collected from microphones, an accelerometer, and gyroscope. The authors segment recordings into 0.4-second intervals, each with a binary label indicating the presence or absence of a cough. We use only the data from the outwardly facing microphone of the recording device.

\subsection{Implementation Details}

All audio is first resampled to 16000Hz and transformed into log mel-spectrograms. If the model underwent pre-training, we normalise the audio using the mean and standard deviation of the pre-training dataset. We evaluate models using 5-Fold Cross-Validation, fine-tuning on the training folds. We adopt the Area Under the Receiver Operating Characteristic Curve (AUROC) as our evaluation metric for its resistance to class imbalance and to align with the metrics used by the dataset authors which all use it. For fine-tuning, we replace the head of each model with a simple classification head appropriate for the task. This maps a sequence level representation from the encoder to output nodes representing each class using a fully connected layer. We explore representing the entire sequence using a classification token or mean pooling each encoded patch.

\renewcommand{\arraystretch}{0.8} 
\begin{table*}[t]
\caption{
Evaluation of Model Performance on downstream Cough Audio Datasets. Compares AUROC(\%) after fine-tuning on target task. Top model results are in bold (best) and underlined (second best). Our proposed model is highlighted.
}
\label{tab:main-results}
\begin{tabular}{l l c c c c c}
\toprule
Model    & 
Pre-training Dataset &
Pre-training Epochs
&
Self-Supervised 
& COVID-19 Detection 
& Cough Detection 
& Wet-or-Dry\\ 
\midrule
ViT      & None                 & - & \xmark          & 56.19              & 97.24           & 60.06                     \\
ViT      & C19-Sounds           & 100 & \xmark          & 59.22              & 95.89           & 58.95                     \\
AST      & None                 & - & \xmark          & 59.79              & 97.26           & 58.23                     \\
AST      & C19-Sounds           & 100 & \xmark          & 60.42              & 94.37           & 55.54                     \\
AST      & Audioset             & 5   &\xmark          & \underline{70.63}  & \textbf{98.73}  & \textbf{78.95}            \\
\rowcolor{lightgray}
CoughViT & C19-Sounds           & 100 & \cmark          & \textbf{73.21}     & \underline{98.25} & \underline{74.95}        \\ 
\bottomrule
\end{tabular}
\end{table*}

\subsection{Models}


We evaluate on all three benchmark tasks our model CoughViT which combines domain-specific and self-supervised pre-training. We explore a variety of configurations and compare the best performing ones to existing methods. 
To evaluate the performance of our pre-training approach, we train the same ViT-Base architecture using more conventional methods. The first involves no pre-training to evaluate the performance gained by pre-training. Second, use a ViT model pre-trained in a supervised manner using COVID-19 Sound's provided annotations (ViT-C19S). There serve as ways to evaluate the effectiveness of domain-specific self-supervised pre-training
Audio Spectrogram Transformer (AST) was the first ViT variant to be adapted to audio data and remains the most popular. Architecture-wise, it differs from CoughViT in two major aspects. Firstly, AST uses overlapping patches when splitting spectrograms while CoughViT uses non-overlapping patches. This results in AST generating a significantly larger number of patches and results in longer training times. Secondly, AST uses a learnable positional embedding to retain positional information while CoughViT uses fixed sinusoidal encodings.
We evaluate on our benchmark tasks three variants of AST: without pre-training, COVID-19 Sounds pre-training (AST-C19S), and Audioset pre-training (AST-Audioset). We contrast the different architectural choices by also evaluating AST without pre-training and with supervised pre-training on COVID-19 Sounds. Finally, we compare our domain-specific pre-training approach against AST-Audioset which is representative of state-of-the-art broad supervised pre-training.

\subsection{Main Results}

\begin{table}[]
\caption{COUGHVID Blind Test Set}
\label{tab:coughvid-test}
\renewcommand{\arraystretch}{0.9}
\begin{tabular}{lc}
\toprule
Model    & AUROC \\ \midrule
Logistic Regression \cite{orlandic2023semi} & \underline{0.59} \\
AST-Audioset      & 0.56  \\
\rowcolor{lightgray}
CoughViT & \textbf{0.71}  \\ \bottomrule
\end{tabular}
\end{table}

\begin{table}[]
\caption{Edge-AI Cough Detection Blind Test Set}
\label{tab:cough-detection-test}
\begin{tabular}{lcc}
\toprule
Model    & Event-based F-1 & Sample-based F-1 \\ \midrule
AST-Audioset      & \textbf{0.71}            & \textbf{0.59}             \\
\rowcolor{lightgray}
CoughViT & \underline{0.69}            & \underline{0.57}            \\ \bottomrule
\end{tabular}
\end{table}

The main experimental results are reported in Table \ref{tab:main-results}. Overall, the best performing models were AST-Audioset and CoughViT. CoughViT performed the best for COVID-19 detection while AST-Audioset performed the best for Cough detection and Wet-or-Dry cough classification.
Self-supervised pre-training for CoughViT improved performance across all benchmark tasks compared to a ViT with no pre-training. Specifically, it increased AUROC(\%) by 17.02 for COVID-19 detection, 1.01 for cough detection, and 14.89 for wet-or-dry classification. This illustrates that patch reconstructions is an effective task for cough representation learning.

\renewcommand{\arraystretch}{0.8} 
\begin{table*}[]
\caption{Ablation Study of CoughViT Configurations. Compares AUROC(\%) after fine-tuning on downstream tasks, with all pre-training conducted on the COVID-19 Sounds dataset. Variations included pre-training strategy and duration, masking ratio, self-attention type for spectrogram reconstruction, and classification method.}
\label{tab:coughvit-ablation}
\small
\begin{tabular}{ccccccccc}
\toprule
Pre-training & 
Self-Supervised & 
\begin{tabular}[c]{@{}c@{}}Pre-training \\ Epochs\end{tabular} & 
\begin{tabular}[c]{@{}c@{}}Masking \\ Ratio \end{tabular} & 
\begin{tabular}[c]{@{}l@{}}Windowed \\ Self-Attention\end{tabular} & 
\begin{tabular}[c]{@{}l@{}}Classification \\ Token\end{tabular} & 
\begin{tabular}[c]{@{}l@{}}COVID-19 \\ Detection\end{tabular} &
\begin{tabular}[c]{@{}l@{}}Cough \\ Detection\end{tabular} &
Wet-or-Dry \\ 
\midrule
\xmark & N/A    & N/A & N/A & N/A    & \xmark & 58.82 & 97.23 & 62.56 \\
\xmark & N/A    & N/A & N/A & N/A    & \cmark & 56.19 & 97.24 & 60.06 \\
\cmark & \xmark & 100 & N/A & N/A    & \xmark & 58.69 & 96.13 & 58.27 \\
\cmark & \xmark & 100 & N/A & N/A    & \cmark & 59.22 & 95.89 & 58.95 \\
\cmark & \cmark & 100 & 0.50& \xmark & \xmark & 69.50 & 97.74 & 70.54 \\
\cmark & \cmark & 100 & 0.50& \xmark & \cmark & 71.95 & 98.05 & 69.50 \\
\cmark & \cmark & 100 & 0.50& \cmark & \xmark & 68.84 & 97.34 & 67.13 \\
\cmark & \cmark & 100 & 0.50& \cmark & \cmark & 67.83 & 97.37 & 58.57 \\
\cmark & \cmark & 100 & 0.75& \xmark & \xmark & \textbf{75.82} & 98.14 & 71.64 \\
\cmark & \cmark & 100 & 0.75& \xmark & \cmark & 73.21 & \textbf{98.25} & \textbf{74.95} \\
\cmark & \cmark & 100 & 0.75& \cmark & \xmark & 69.95 & 97.76 & 68.40 \\
\cmark & \cmark & 100 & 0.75& \cmark & \cmark & 69.39 & 97.74 & 68.38 \\
\cmark & \cmark & 200 & 0.75& \xmark & \xmark & 74.40 & 98.08 & 71.06 \\
\cmark & \cmark & 200 & 0.75& \xmark & \cmark & \underline{74.73} & \underline{98.23} & \underline{73.90} \\ 
\bottomrule
\end{tabular}
\end{table*}

AST-Audioset's performance underscores the efficacy of general audio pre-training. This style of pre-training also improves performance for all tasks compared to no pre-training. It increases AUROC(\%) by 10.84 for COVID-19 detection, 1.47 for cough detection, and 20.72 for wet-or-dry classification. This suggests that leveraging off-the-shelf models for feature extraction or fine-tuning is an effective and practical strategy for cough audio modelling.

Comparing these two top performing models: CoughViT outperforms AST-Audioset in COVID-19 detection by 2.58 but is outperformed in cough detection and wet-or-dry classification by 0.48 and 4.0 respectively. This result highlights the advantages of pre-training on in-domain data. Despite being trained on a smaller unlabelled dataset CoughViT competes closely with AST-Audioset, which benefited from an extensive labelled pre-training dataset.

We found that supervised pre-training on COVID-19 Sounds showed minimal benefit for both AST and ViT architectures. For COVID-19 Detection, this pre-training offered a slight performance improvement. However, it negatively impacted performance for other tasks compared to no pre-training. This aligns with the observation that supervised pre-training leads to overfitting on the pre-training task. The self-reported nature of the labels for this pre-training task may have compounded these issues.

To compare the two model architectures we highlight two scenarios: without pre-training and with supervised pre-training using COVID-19 Sounds. Without pre-training, ViT exhibits a slight advantage over AST in wet-or-dry classification but falls slightly short in the other two tasks. When supervised pre-training is applied, ViT-C19S shows better performance in both cough detection and wet-or-dry classification, yet lags in COVID-19 Detection compared to AST-C19S. These differences are minor indicating that the architectural variations do not significantly impact performance. As explored previously AST encodes a larger number of patches due to its employment of overlapping patches. These findings imply that employing non-overlapping patches can yield comparable outcomes with greater computational efficiency.

\subsection{Blind Test Sets}

Where available, we submit our top performing models, CoughViT and AST-Audioset, for evaluation on blind test sets withheld by the dataset authors. These were the COUGHVID Dataset \cite{orlandic2021coughvid} and Edge-AI Cough Detection Dataset \cite{orlandic2023multimodal}. For both tasks we retrain the models on the entire public dataset, taking the average of the best performing number of epochs for each fold from cross validation. We provide results in the format given to us.
Using the blind COUGHVID test set, we perform a evaluation for wet-or-dry cough classification for annotations from expert 4. CoughViT significantly outperformed both AST-Audioset and a logistic regression model previously evaluated \cite{orlandic2023semi}.
The blind test set of the Edge-AI Cough Detection Dataset evaluates models on Cough Segmentation. To adapt our binary Cough Detection model for this task, we employ a sliding window technique across the audio samples. 
The sliding windows move through the audio in fixed steps of 0.01 seconds. When a cough is detected, its start and end times are set to the edges of the window. If another cough is detected in a subsequent overlapping window, the end time is revised to the start of the subsequent cough.

Two approaches are used for evaluation: sampled-based and event-based scoring. The sampled-based approach segments the audio into uniform intervals. The ground truth timestamps and model-generated timestamps are used to label the presence of a cough in each interval. The metric is calculated by comparing interval classifications of the model against the ground truth. Event-based scoring instead determines if a cough was correctly segmented by using the overlap between the ground truth and model generated timestamps. For both approaches, AST slightly outperforms CoughViT using the F-1 metric. At the time of evaluation, ours were the first models to be evaluated on this blind test set.

\subsection{CoughViT Ablation}
To further validate the effectiveness of the proposed model, we conduct experiments with different configurations. 
Table \ref{tab:coughvit-ablation} summarises the performance of different configurations. Specifically, the following aspects have been investigated.
(1) \textbf{Sample Level Representation:} Transformers take sequences as input, a sequence of patches in the case of Vision Transformers, and use an encoder to embed each item in the sequence. However, for tasks such as classification a representation for the entire sample is required. Typically this is done by pre-pending a classification token to the sequence which acts as a representation of the entire input. An alternative method is to take the mean of each encoded patch. We explore both strategies throughout the experiments and find minor differences for all tasks. This suggests that mean pooling is a valid approach and may slightly simplify the model architecture.
(2) \textbf{Masking Ratio:} We also explore the amount of masking applied during self-supervised pre-training. Too much masking and the Image Reconstruction task becomes too difficult, however, too little and the model doesn't learn enough knowledge about the domain. We find that models that used 50\% masking during pre-training still perform well on downstream tasks, they are consistently outperformed by the ones that used 75\% masking.
(3) \textbf{Decoder Attention Mechanism:} We investigated the use of windowed self-attention in the CoughViT decoder, hypothesising that cough audio may not have extensive long-range dependencies. Contrary to our expectations, this approach resulted in diminished performance compared to the standard self-attention mechanism.
(4) \textbf{Pre-training Epochs:} For pragmatic reasons, we cut off pre-training at 100 epochs, although pre-training validation loss for the self-supervised task was still dropping. To explore the effect of further pre-training we used the best performing configuration of CoughViT and continued pre-training for an additional 100 epochs. We find this version performs slightly better for downstream COVID-19 Detection, similarly for Cough Detection, and slightly worse for cough wet-or-dry classification. This minimal change in performance suggests the model has learnt most of the valuable information from the self-supervised task by 100 epochs.

\section{Conclusion}

In this study, we propose a novel framework to learn general-purpose cough representations from audio signals. Through domain-specific pre-training, our representations enhance classifier performance for data limited tasks. We also employ self-supervised learning on unlabelled datasets to address label scarcity challenges. To our knowledge, we are the first to propose a general-purpose cough representation learning framework. Additionally, this study appears to be the first to demonstrate the Vision Transformer architecture's effectiveness in cough audio modelling. Our experimental results show that our learned representations match or exceed the state-of-the-art in enhancing performance on diagnostically important tasks.

A lack of variety in publicly available cough audio datasets has limited our evaluation of the tasks explored in this paper. Evaluating our approach across a broader range of respiratory conditions represents a critical next step in research. Furthermore, we note the recent release of the largest cough audio dataset to date, the "UK COVID-19 Vocal Audio Dataset" \cite{budd2022large}, which includes clinically validated annotations. This dataset is a promising resource for further research on domain-specific pre-training or as a benchmark. Finally, we propose that ensembles of respiratory disease classifiers enhanced by CoughViT feature representations offer significant potential for advancing research in AI-based differential diagnosis.

\begin{acks}
This research is supported by the Australian Commonwealth Scientific and Industrial Research Organisation and the United States National Science Foundation under Grants No.2302968, No.2302969, and No.2302970, titled ``Collaborative Research: NSF CSIRO: HCC: Small: Understanding Bias in AI Models for the Prediction of Infectious Disease Spread''.
We would like to thank the support of the National Computational Infrastructure and the ARC Center of Excellence for Automated Decision Making and Society (CE200100005).

\end{acks}
\bibliographystyle{ACM-Reference-Format}
\bibliography{main}


\end{document}